\documentclass[a4paper]{jpconf}
\usepackage{graphicx}
\begin{document}
\title{A Wannier orbital based method for resonant inelastic x-ray scattering simulation}

\author{Chunjing Jia}

\address{Stanford Institute for Materials and Energy Sciences, SLAC National Accelerator Laboratory and Stanford University, Menlo Park, California 94025, USA}

\ead{chunjing@stanford.edu}

\begin{abstract}
We report one algorithm for simulating oxygen {\it K}-edge RIXS for weakly correlated systems, using maximally localized Wannier functions as the basis set. The $N$-electron wavefunction is formulated using the single Slater determinant, and the many-body effect is treated explicitly at the dipole matrix element level. The simulated result of oxygen {\it K}-edge RIXS for solid state Li$_2$CO$_3$ matches well with the experimental data. Besides being efficient and reasonably accurate, this algorithm also shows potential to extent to more complex RIXS problems.
\end{abstract}

\section{Introduction} 
Since its discovery by Wilhelm Roentgen more than a hundred years ago, x-ray has not only improved the quality of life of human beings enormously via chemical imaging and radio therapy etc, but also provided scientists a powerful tool to directly investigate the microscopic scale of materials. Use transition-metal oxide as an example: x-rays with high photon energy above 5-10 keV, or usually called hard x-rays, can been tuned to coincide with the energy difference between a certain transition-metal $1s$ core-level and valence, thus selectively exciting the electronic environment of the transition metal in the system. x-rays with lower photon energies, or soft x-rays, can be used to excite transition-metal $2p$ electrons or oxygen $1s$ electrons in transition metal oxides. The energy and edge selectivity of x-ray spectroscopies and their large cross-section (or no requirement for using large sample size) are among the main advantages of x-ray spectroscopic technique. Thanks to the recent development of high throughput high resolution synchrotron light sources and detection instruments, x-ray spectroscopies have become increasingly important in many fields.  In physics, x-ray absorption spectroscopy (XAS) and resonant inelastic x-ray scattering (RIXS) have been extensively employed for understanding electronic structures generally and collective excitations specifically in strongly correlated systems~\cite{RIXSReview, Tacon2011, Dean2013, Huang2016}. In chemistry, x-ray spectroscopy have been used as a fingerprint of electronic structure to look into static or transient states of matter.~\cite{Kelly2015} Recently, XAS and RIXS have also been used to search for better battery materials by investigating the redox process in the materials science community~\cite{Gent2017, Yang2018}. 

While x-ray spectroscopies have been extensively and successfully used in various of fields, theoretical understanding of them remains a great challenge. The reason is that we lack of an universal theoretical scheme to treat different systems (For certain systems correlation effect or localized physics is important, while for other systems band dispersion matters more. What's worse is that for many systems both effects are important!) for different edges (transition-metal {\it K}, {\it L}, {\it M} edge or oxygen {\it K} edge etc).  One of the main theoretical difficulties lies in the choice of basis, as how the wavefunction is formulated is very different between weakly and strongly correlated electronic systems. See Table 1 for a comparison between the single-particle and the many-body methods of construcing Hilbert space. To treat strongly correlated electronic systems one has to use the full many-body basis as the starting point to construct the Hamiltonian matrix, and the size of the Hilbert space scales exponentially with the number of available single-particle states (which could be the number of orbital for atomic system, the number of real space lattice size for single-band Hubbard model, or the combination of both for a realistic multi-orbital/multi-site system). This induces strong limitation on the largest systems one can solve, and usually band dispersion which comes along with the inclusion of hoppings in the intermediate and long range can not be addressed. On the other hand, for weakly correlated electronic systems, the many-body electronic wavefunction can be expressed as the single Slater determinant of the single-particle states, thus the size of the Hilbert space scales only linearly with the number of single-particle states. Using this basis set and construction we can easily address the band dispersion by using a supercell with hundreds of atoms. However, part of the many-body effect is missing ``institutionally" as the single Slater determinant wavefunction construction can not catch the complex response of the rest $N-1$ electrons when one electron is being disturbed.  For certain systems, such as strongly correlated electronic systems, this missing part can be the dominant and not including it is fatal in describing the characteristic of the material. In a word, with current computing power and current algorithms, it has not been possible to address both the strongly correlation effect and the intermediate/long range hoppings properly at the same time.

\begin{table}
\begin{center}
\includegraphics[width=30pc]{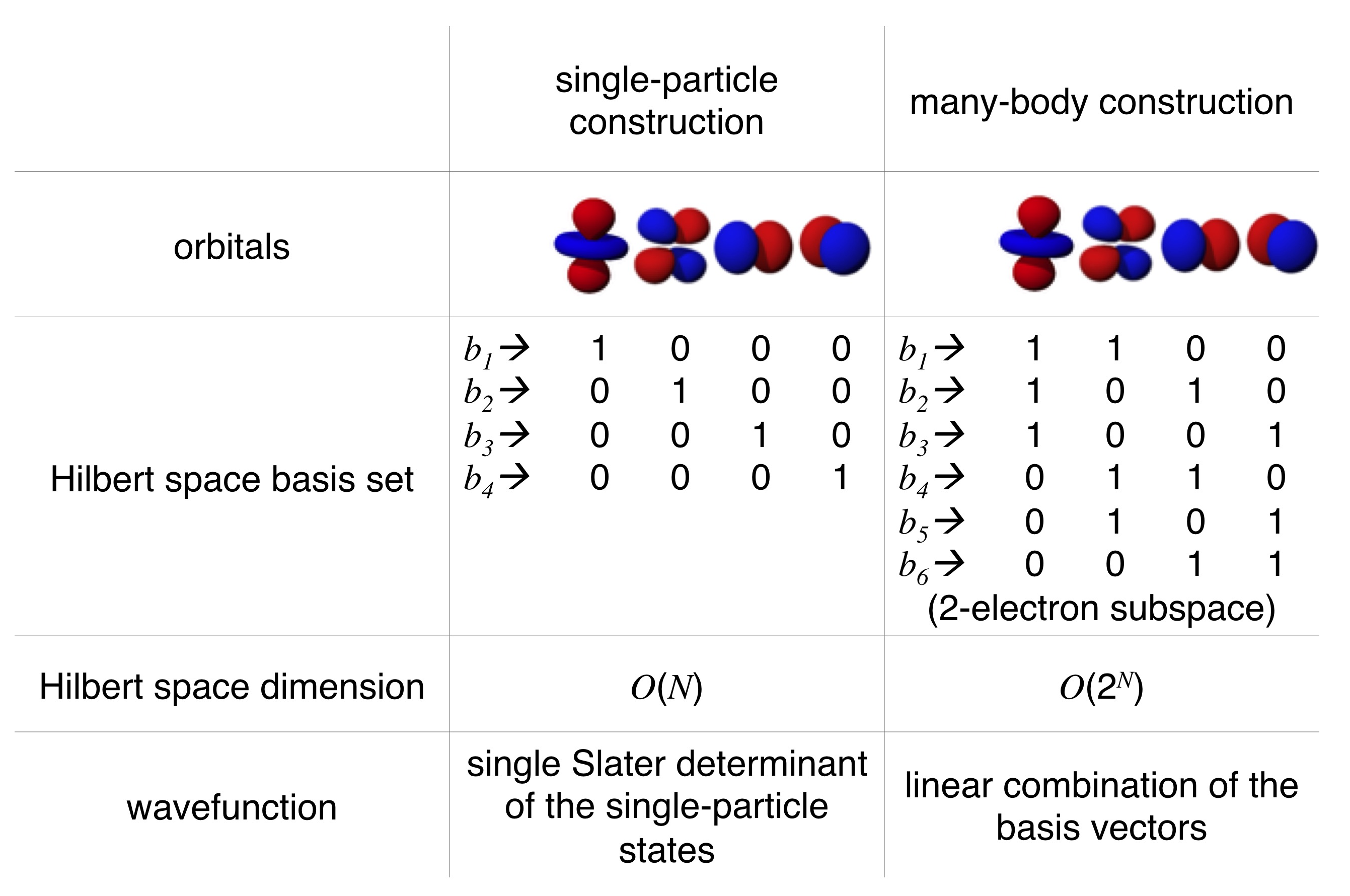}
\end{center}
\caption{\label{label} Comparison of single-particle and many-body constructions for a spinless four-orbital system.}
\end{table}

Instead of directly tackling this hard problem of finding a universal theory for understanding x-ray spectroscopies of both strongly and weakly correlated systems/theories, scientists in the theoretical x-ray spectroscopy community so far have taken another route: looking at each problem carefully and developing a number of different methods each suitable for a specific set of problems.  {\it ab initio} based methods such as FEFF~\cite{FEFF} and OCEAN~\cite{OCEAN} have been developed to target weakly correlated systems and oxygen {\it K}-edge spectra. Many-body techniques such as exact diagonalization have been used for strongly correlated systems and transition-metal {\it K}/{\it L}-edges.~\cite{JiaKedgeRIXS, JiaLedgeRIXS1, JiaLedgeRIXS2}

In this study, we will report one newly developed method that adds to the existing algorithm ``zoo" for x-ray spectroscopy simulations. We target for oxygen {\it K}-edge for weakly correlated systems or at least ``not so" strongly correlated systems. We will use the single Slater determinant to construct the $N$ electron wavefunction, and treat the many-body effect explicitly on the matrix element level. The main advantage of this method is that instead of using plane-wave basis that have been usually used in the {\it ab initio} methods for spectroscopy simulation for solid state materials, we use Wannier orbitals as the single-particle basis. As shown in Table 1, for the single-particle construction there is a one-to-one correspondence between each building block (Wannier orbital in our method or plane-wave for solid state DFT simulations) and Hilbert space basis vector, thus we will use the two terms, Wannier orbitals and basis set, interchangeably sometimes in the discussion our algorithm. Basis using Wannier orbitals can be demonstrated to be the most compact basis set for valence electrons, thus can largely accelerate the algorithm and enhance the numerical accuracy.  Moreover, since Wannier orbitals are very close to atomic orbitals in the solid state systems, by construction it is easier to extend the current algorithm to include correlation effect. Besides the above advantages, what scientists care most about a numerical simulation is always how good the result is. For this reason, we will demonstrate the effectiveness of this algorithm by showing the numerical results of RIXS and its comparison with experiment for an example material. The numerical simulation shows reasonably good result and even catches the fine details of the experimental data for the example material Li$_2$CO$_3$ at the oxygen {\it K}-edge.

The rest of the manuscript is organized as the following: In section 2, we will talk about the methods by showing the derivation of the algorithm for XAS and RIXS cross-sections. In section 3, we will show the numerical results of RIXS and its comparison with experimental data for Li$_2$CO$_3$ at the oxygen {\it K}-edge. In section 4, we will conclude this paper with summaries and discussions. 

\section{Methods}

\subsection{Hamiltonians}

The tight-binding Hamiltonian of the multi-orbital system can be expressed as:
\begin{equation}
H_0 = \sum_{i, j, \alpha_i, \alpha_j, \sigma} t_{i, j, \alpha_i, \alpha_j} c^\dagger_{i,\alpha_i,\sigma} c_{j,\alpha_j,\sigma}
\end{equation}
where index $i$/$j$ is the atom index, $\alpha_i$ represents the orbital index at atom $i$, $\sigma$ represents spin. $t_{i, j, \alpha_i, \alpha_j}$ represents the hopping between orbital $\alpha_i$ at atom $i$ and orbital $\alpha_j$ at atom $j$, and especially when $i = j$ and $\alpha_i = \alpha_j$,  $t_{i, i, \alpha_i, \alpha_i}$ represents the site energy. This Hamiltonian applies to both the RIXS/XAS initial state and RIXS final state without core-hole.

For the electronic state with a core-hole (RIXS intermediate state or XAS final state) at site $i_0$, suppose the only interaction term brought in by the core-hole is the monopole term of the Coulomb interaction, then the Hamiltonian with one core-hole can be written as:
\begin{equation}
H_i = \sum_{i, j, \alpha_i, \alpha_j, \sigma} t_{i, j, \alpha_i, \alpha_j} c^\dagger_{i,\alpha_i,\sigma} c_{j,\alpha_j,\sigma} - \sum_{\alpha_{i_0}, \sigma} U_c n_{i_0, \alpha_{i_0}, \sigma} n^h_{i_0, \sigma^\prime} + \epsilon({O_{h}}) n^h_{i_0, \sigma^\prime}
\end{equation}
where $n^h_{i_0, \sigma^\prime}$ is the number operator for core-hole with spin $\sigma^\prime$ at site $i_0$, ${\epsilon(O_{h})}$ is the core-hole site energy, $U_c$ is the core-hole potential and it only applies to the orbitals on the same site $i_0$, $n_{i_0, \alpha_{i_0}, \sigma}$ is the electron number operator for orbital $\alpha_{i_0}$ at site $i_0$.  Please note that systems with higher order interactions of the core-hole such as multiplet interactions can not be described by the above Hamiltonian. For this reason, the Wannier orbital based RIXS algorithm we propose here is suitable for simulating XAS and RIXS at oxygen {\it K}-edge (sulfur {\it K}-edge etc), but not transition-metal {\it L}-edge where the $3d-2p$ multiplet interaction on the transition metal atom is usually important. 
 
There are a number of ways to obtain the tight-binding model Hamiltonian for a material specific problem. What we use here is the maximally localized Wannier functions method implemented in Wannier90~\cite{Wannier90}, by projecting the {\it ab initio} density functional theory (DFT) results onto the maximally localized Wannier basis set to obtain the Hamiltonian of our interest. Compared with directly fitting the DFT bandstructure, each basis obtained in the above Wannier method has a physical meaning (They are Wannier orbitals! They are also maximally localized to repress long range hoppings.) thus it preserves the symmetries of the problem by nature.

\subsection{XAS}
The general XAS formalism in the single Slater determinant construction has been discussed previously by Liang {\it et al}~\cite{Liang2017}. Here we will firstly rederive them for XAS, and then move on to the derivation of RIXS.

The initial state of the $N+1$ electron wavefunction for XAS is the ground state of the electronic system of $N$ valence electrons plus one core-electron. (We only consider one core electron in the initial state or one core-hole state in the final state of XAS.) This $N+1$ electron wavefunction can be expressed as $ | \Phi_0 \rangle$ = $\prod_{v=1}^{N} b_v h | 0 \rangle$, where $| 0 \rangle$ is the empty state,  $h$ is the core-electron creation operator (or core-hole annihilation operator), $b_v$ annihilates a valence hole (or creates a valence electron) with single particle index $v$. $ | \Phi_0 \rangle$ should be understood as the single Slater determinant of the $N+1$ one-electron wavefunction.

The final state $N+1$ electron wavefunction for XAS can be presented in the following:
\begin{equation}
| \Phi_i \rangle = \prod_{\mu=1}^{N+1} \tilde{\phi}^\dag_{i_\mu} \prod_{v=1}^{N} b^\dag_v h^\dag | \Phi_0 \rangle \\
= \prod_{\mu=1}^{N+1} ( \sum_c \xi_{{i_\mu}, c} a^\dag_c + \sum_v \xi_{i_\mu, v} b_v)  \prod_{v=1}^{N} b^\dag_v h^\dag | \Phi_0 \rangle
\end{equation}
where $| \Phi_0 \rangle$ is the initial state $N+1$ electron wavefunction, $a^\dagger_c$ creates an electron with index $c$ in the conduction band, and $\prod_{v=1}^{N} b^\dag_v h^\dag | \Phi_0 \rangle$ represents the state with empty electrons in the valence and one core-hole. $\tilde{\phi}^\dag_{i_\mu}$ creates an electron with state index $i_\mu$ for the XAS final  state Hamiltonian with core-hole, and $\tilde{\phi}^\dag_{i_\mu}$ is connected with the initial state single particle eigenstates without core-hole by the relation:  $\tilde{\phi}^\dag_{i_\mu} = \sum_c \xi_{{i_\mu}, c} a^\dag_c + \sum_v \xi_{i_\mu, v} b_v$.

XAS cross-section can be expressed in the Fermi's Golden rule:
\begin{equation} 
I(\omega) = \sum_{i} | \langle \Phi_i | \epsilon \cdot \mathbf{R} | \Phi_0 \rangle | ^2 \delta (E_i - E_0 -  \omega) 
\end{equation} 
Here $| \Phi_i \rangle$ is the $N+1$ electron final state wavefunction and $| \Phi_0 \rangle$ is the $N+1$ electron initial state wavefunction. $E_0$ is the sum of the $N$ lowest single-particle energies in the valence $E_0 = \sum_{v=1}^N e_v$, and   $E_i$ is the sum of the single-particle energies for final state $i$:  $E_i = \sum_{\mu=1}^{N+1} \tilde{e}_{i_\mu} + e_h$. As we have stated in the introduction, we will treat them in the Slater determinant of the single particle basis, and consider the many-body effect by treating the matrix elements explicitly. The matrix element can be expressed in the single particle basis without core-hole: 

\begin{equation} 
 \langle \Phi_i | \epsilon \cdot \mathbf{R} | \Phi_0 \rangle =  \sum_c (A^i_c)^* \langle \phi_c | \epsilon \cdot r | \phi_h \rangle 
\end{equation} 
where $| \phi_h \rangle$ is the single-particle wavefunction of the core-electron and $| \phi_c \rangle$ is the single-particle wavefunction of a conduction state with index $c$. $\langle \phi_c | \epsilon \cdot r | \phi_h \rangle$ shows the dipole of one electron excited from the core up to a conduction state with index $c$; the coefficient $A^i_c$ shows the overlap of the ($N+1$)-electron wavefunction between the initial state and the final state, and can be expressed as:

\begin{equation}
A^i_c = det
\left( \begin{array}{ccccc}
\xi_{i_1, v=1} 	& \xi_{i_1, v=2} 	& \ldots 	& \xi_{i_1, v=N} 	& \xi_{i_1, c} \\
\xi_{i_2, v=1} 	& \xi_{i_2, v=2} 	& \ldots 	& \xi_{i_2, v=N} 	& \xi_{i_2, c} \\
\vdots 		& 			& \ddots 	&			& \vdots	\\
\xi_{i_{N+1}, v=1} 	& \xi_{i_{N+1}, v=2} & \ldots 	& \xi_{i_{N+1}, v=N} & \xi_{i_{N+1}, c} \\
\end{array} \right)
\end{equation}

\subsection{RIXS}
 
RIXS cross-section is expressed in Kramer's Heissenberg formular~\cite{RIXSReview}:
\begin{equation} I(\omega_{in}, \omega_{loss}) = \sum_{f} | \sum_{i} \langle  \Phi_f | \epsilon \cdot \mathbf{R} | \Phi_i \rangle  \frac{1}{E_i - E_0 - \omega_{in} - i\Gamma} \langle  \Phi_i | \epsilon \cdot \mathbf{R} | \Phi_0 \rangle | ^2 \delta (E_f - E_0 - \omega_{loss})  \end{equation} 

The RIXS intermediate state wavefunction $| \Phi_i \rangle$ has the same expression as the XAS final state wavefunction. The RIXS final state $N+1$ electron wavefunction can be expressed as:
\begin{equation}
| \Phi_f \rangle = a^\dag_c b^\dag_v | \Phi_0 \rangle + a^\dag_{c1} a^\dag_{c2} b^\dag_{v1} b^\dag_{v2} | \Phi_0 \rangle + \ldots \approx a^\dag_c b^\dag_v | \Phi_0 \rangle
\end{equation}

This approximation only keeps the first term (or a single electron-hole pair) in the final states. For intermediate state $| \Phi_i \rangle = \prod_{\mu=1}^{N+1} ( \sum_c \xi_{{\mu}, c} a^\dag_c + \sum_v \xi_{\mu, v} b_v)  \prod_{v=1}^{N} b^\dag_v h^\dag | \Phi_0 \rangle$ and final state $| \Phi_{f=(v,c)} \rangle = a^\dag_c b^\dag_v | \Phi_0 \rangle$, the matrix element can be expressed as:
\begin{equation} 
\langle  \Phi_{f=(v,c)} | \epsilon \cdot \mathbf{R} | \Phi_i \rangle =  A^i_c \langle \phi_h | \epsilon \cdot r | \phi_v \rangle 
\end{equation}

By expanding the matrix elements with single-particle wavefunction overlaps, the above RIXS cross-section can be expressed as:

\begin{equation} 
I(\omega_{in}, \omega_{loss}) \\
= \sum_{f=(v,c)} | \sum_{i} (A^i_c) \langle \phi_h | \epsilon_{out} \cdot r | \phi_v \rangle  \frac{1}{E_i - E_0 - \omega_{in} - i\Gamma} (A^i_c)^* \langle \phi_c | \epsilon_{in} \cdot r | \phi_h \rangle | ^2 \delta (E_f - E_0 - \omega_{loss}) 
\end{equation} 
where the final state energy $E_{f=(v,c)} = \sum_{v'=1}^N e_{v'} - e_v + e_c$.

The above formular is the general expression of RIXS cross-section for one electron-hole pair excitations in the final state. In the following calculation, we will take one more approximation: for XAS final state or RIXS intermediate state $| \Phi_i \rangle$, we only consider those many-body states with $N$ lowest single-particle states occupied (the remaining one electron can occupy any unfilled single-particle state at arbitrary energy):
\begin{equation}
| \Phi_i \rangle = \tilde{\phi}^\dag_{\mu} \prod_{l=1}^{N} \tilde{\phi}^\dag_{l} \prod_{v=1}^{N} b^\dag_v h^\dag | \Phi_0 \rangle.
\end{equation}

For large band gap insulator, intermediate state with one or more electron-hole pairs corresponds to much larger excitation energy.~\cite{Liang2017} As we will show in the next section, the above approximation gives reasonable result for the incoming energy range near the excitation edge.

\section{Results}

In this section, we show the numerical simulation of O {\it K}-edge RIXS spectra for an example material Li$_2$CO$_3$. Experimental RIXS spectra of Li$_2$CO$_3$ at the oxygen {\it K}-edge have been measured by Zhuo and his colleagues~\cite{YangExperiment}. This material is interesting for material scientists and physicists because of its connection with Li-ion battery materials to investigate the possible explanation of anionic redox. The crystal structure of Li$_2$CO$_3$ has been shown at Figure 1(a) and (b). Each primitive cell has four Li$^+$ and two CO$_3^{2-}$. Solid state  Li$_2$CO$_3$ is an insulator with band gap $\sim$ 5eV, with its band structure and partial density of states shown at Figure 1(c). The above bandstructure and density of states were calculated with the density functional theory package {\it Quantum Espresso}~\cite{QE} using PBE exchange-correlation functional, norm-conserving pseudopotentials, and a $4 \times 4 \times 4$ momentum sampling.

\begin{figure}
\begin{center}
\includegraphics[width=37pc]{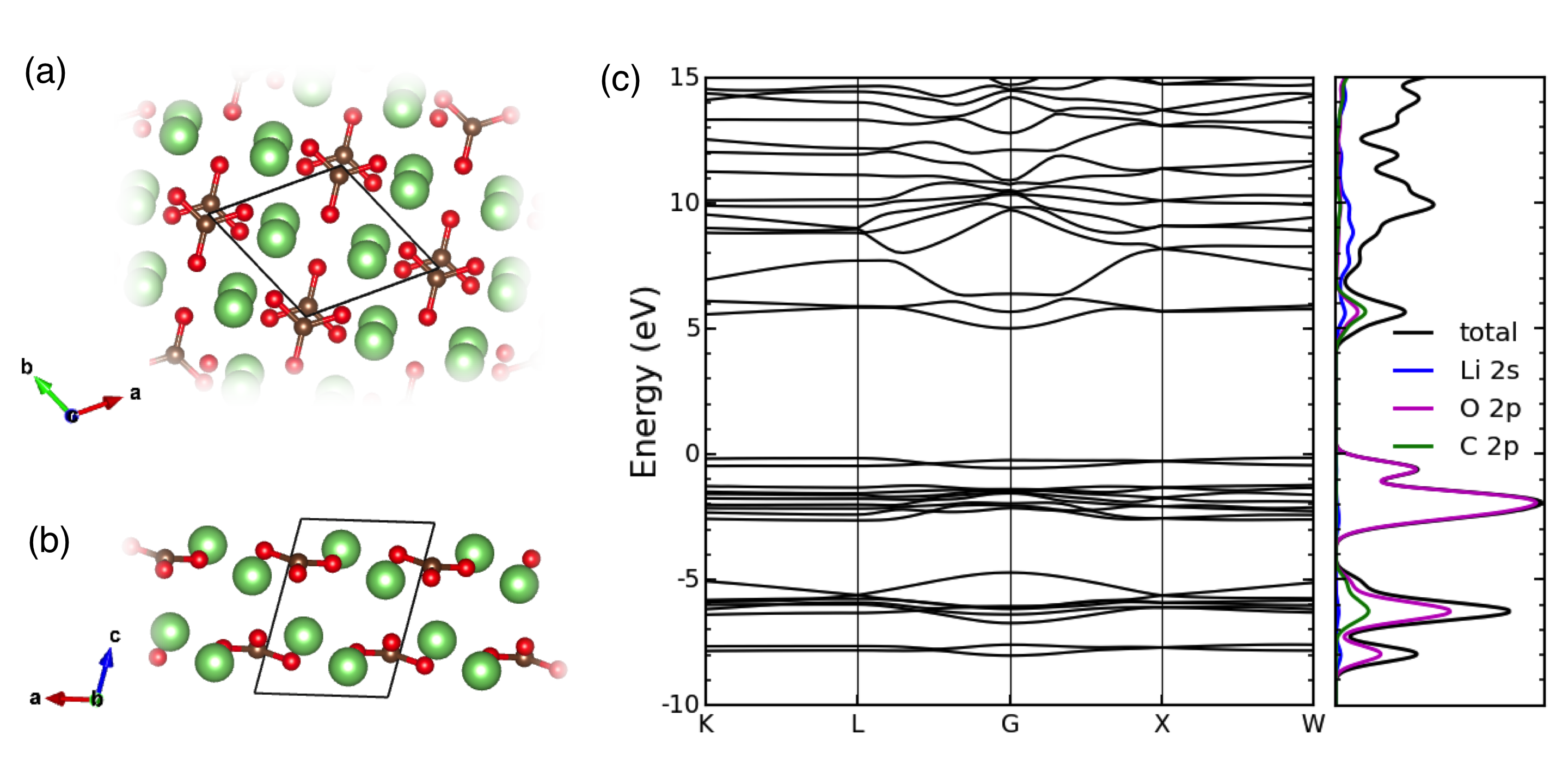}
\end{center}
\caption{\label{label} (a) Topview and (b) sideview of Li$_2$CO$_3$ crystal structure. Green, brown and red spheres represent Li, C and O atoms respectively. Black frame shows the single primitive cell boundary. (c) DFT calculated band structure and partial density of states of Li$_2$CO$_3$. Energy $0$ represents fermi level $E_f$.}
\end{figure}

The Wannier downfolding using Wannier90~\cite{Wannier90} was then implemented with the initial projection onto carbon $pz$ orbitals, oxygen $px$/$py$/$pz$ orbitals, Li $s$ orbitals and ten $s$-like orbitals at random positions (a total 34-orbital model). Oxygen {\it K}-edge RIXS was calculated using Eq. (10) with a $4 \times 4 \times 4$ supercell. Both the incoming and outgoing x-ray photons were taken to be non-polarized. We used core-hole lifetime broadening $\Gamma = 0.25$eV, core-hole potential $U_c = 3.5$eV and $\epsilon(O_{1s}) = 534$eV (relative to $E_f$) in the calculation.

\begin{figure}
\begin{center}
\includegraphics[width=25pc]{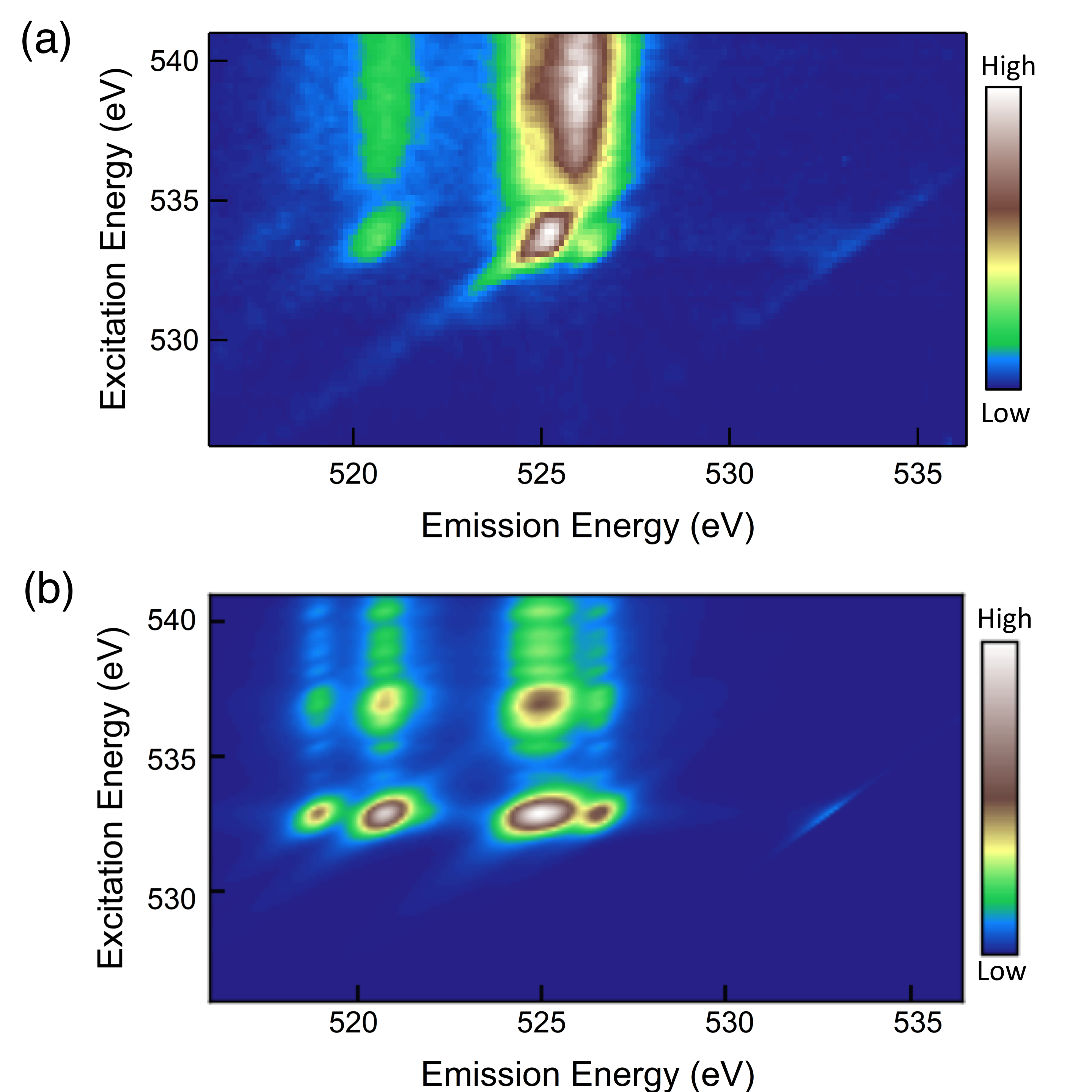}
\end{center}
\caption{\label{label} (a) Experimental O {\it K}-edge RIXS of Li$_2$CO$_3$, reprint from Ref.\cite{YangExperiment}. (b) Theoretical O {\it K}-edge RIXS spectra of Li$_2$CO$_3$ using our method.}
\end{figure}

In Figure 2, the experimental RIXS data shows three main emission features at x-ray emission energies 519eV, 520.5eV and 526eV respectively. Our theoretical simulation reproduced all the three emission features at 519eV, 520.5eV and 525eV respectively. In simulation we see another emission feature at 526.5eV, which is not seen in the experiment. A probable explanation is that in experiment the 526.5eV emission has merged into the main 526eV emission. Simulation with a larger core-hole lifetime broadening $\Gamma = 0.5$eV also reports merging of these two emissions. To understand the nature of these excitations, we can make connection with the calculated partial density of state as shown in Figure 1(c). These four calculated emission lines at 519eV, 520.5eV, 525eV and 526.5eV correspond to the energy it releases for the recombination of valence electrons of oxygen $2p$ orbital content from $\sim$ -8eV, -6eV, -2eV and -0.5eV back to the $1s$ core level. 

At incoming excitation energy $\sim 533eV$, experimental RIXS spectra show four Raman-like (with a finite energy loss) features at emission energies at $\sim$ 518eV, 520.5eV, 525eV, and a sharp peak at 526.5eV on the right. Our numerical simulation reports four Raman-like features at $\sim 533eV$ excitation energy, and their emission energies at $\sim$ 518.5eV, 520.5eV, 525eV, and 526.5eV. The $\sim 533eV$ incoming x-ray energy corresponds to the energy it takes for an electron to be excited from a $1s$ core to unfilled electronic states with O $2p$ orbital content at 5 - 6.5eV, as shown in Figure 1(c). The energy loss of these four Raman-like features corresponds to the energy of the final state with one electron-hole pair, where the electron resides at the conduction band with oxygen $2p$ orbital content and the hole resides at one of the four occupied oxygen $2p$ electronic states. For both the emission lines and the Raman features, some of the calculated energies do not line up perfectly with the experimental data, mainly because of not including the valence Coulomb interaction and excitonic effect in the calculation. However, overall, our numerically simulated Li$_2$CO$_3$ RIXS spectra have well reproduced all the features in experimental data with high resolution.

\section{Conclusions and discussions}

There are several points I would like to discuss about this algorithm. (1) This algorithm is ``almost" {\it ab initio}, since the only adjustable parameters are $U_c$ and $\epsilon(O_{1s})$, both of which can be potentially evaluated in DFT calculations using oxygen pseudopotential with a $1s$ core-hole. In our Li$_2$CO$_3$ calculation, we take $\epsilon(O_{1s})$ to be adjustable to fit with the experimental energy scale. The spectra weight is only weakly dependent on the value of $U_c$. 
(2) At higher incoming energies ($>$ 537eV) where the emission lines dominate, our algorithm underestimates the spectral weight, compared with the experiment. The reason is that the usage of the approximation in Eq.(11) does not include some of the high energy ($>$ absorption edge + band gap) configurations in the intermediate state.~\cite{Liang2017} 
Energy scale close to the absorption edge is what people usually care most for RIXS spectra, since it captures rich information about the ground, intermediate and final states of the RIXS process. The current approximation works well for simulating RIXS spectra close to the absorption edge for large band gap insulator, and it is quite efficient computationally. For semiconductor or metal, the approximation can be changed to allow one or more electron-hole pairs in RIXS intermediate state. In other words, we may need to go beyond approximations as stated in Eq.(11) for these cases. (3) The current theory does not take the Coulomb interaction and excitonic effect into consideration. However, in the Wannier orbital basis set, we can possibly extend the current algorithm by including the Coulomb interaction (both on-site and in the short/intermediate range) to address these correlation effects. (4) Since Wannier orbital basis set is the most compact complete basis set for valence electrons, this algorithm is very efficient in terms of both time and space computational complexity. 

To summarize, we have derived the Kramer's Heissenberg formula of RIXS cross-section by addressing the many-body effect at the matrix element level. The simulated result of oxygen {\it K}-edge RIXS for solid state Li$_2$CO$_3$ matches the experimental data. Besides being efficient and reasonably accurate, this algorithm also shows potential to extent to more complex RIXS problems. 

\ack The author thanks Thomas Devereaux and Brian Moritz for enormous support and help. The author thanks Wanli Yang for extensive discussions about the connection between theory and experiment. The author thanks Ilkyu Lee and Yuan Chen for insightful discussions. The author also acknowledges the fruitful discussions with Yufeng Liang, whose previous studies and strong passion in physics inspire the author a lot. This work was supported at SLAC and Stanford University by the U.S. Department of Energy, Office of Basic Energy Sciences, Division of Materials Sciences and Engineering, under Contract No. DE-AC02-76SF00515.


\section*{References}
\begin{thebibliography}{9}

\bibitem{RIXSReview} Ament, L. J. P., van Veenendaal, M., Devereaux, T. P., Hill, J. P. and van den Brink, J. Resonant inelastic x-ray scattering studies of elementary excitations., 2011, {\it Rev. Mod. Phys.} {\bf83}, 705?767

\bibitem{Tacon2011} M. Le Tacon, G. Ghiringhelli, J. Chaloupka, M. Moretti Sala, V. Hinkov, M. W. Haverkort, M. Minola, M. Bakr, K. J. Zhou, S. Blanco-Canosa, C. Monney, Y. T. Song, G. L. Sun, C. T. Lin, G. M. De Luca, M. Salluzzo, G. Khaliullin, T. Schmitt, L. Braicovich and B. Keimer, Intense paramagnon excitations in a large family of high-temperature superconductors, 2011, {\it Nature Physics}, {\bf 7}, 725

\bibitem{Dean2013} M. P. M. Dean, G. Dellea, R. S. Springell, F. Yakhou-Harris, K. Kummer, N. B. Brookes, X. Liu, Y.-J. Sun, J. Strle, T. Schmitt, L. Braicovich, G. Ghiringhelli, I. Bozovic, J. P. Hill, Persistence of magnetic excitations in La$_{2?x}$Sr$_x$CuO$_4$ from the undoped insulator to the heavily overdoped non-superconducting metal, 2013, {\it Nature Materials}, {\bf 12}, 1019

\bibitem{Huang2016} HY Huang, CJ Jia, ZY Chen, K Wohlfeld, B Moritz, TP Devereaux, WB Wu, J Okamoto, WS Lee, M Hashimoto, Y He, ZX Shen, Y Yoshida, H Eisaki, CY Mou, CT Chen, DJ Huang, Raman and fluorescence characteristics of resonant inelastic x-ray scattering from doped superconducting cuprates, 2016, {\it Scientific Reports}, {\bf 6}, 19657

\bibitem{Kelly2015} Ph Wernet, Kristjan Kunnus, Ida Josefsson, Ivan Rajkovic, Wilson Quevedo, Martin Beye, Simon Schreck, Sebastian Grübel, Mirko Scholz, Dennis Nordlund, Wenkai Zhang, Robert W Hartsock, William F Schlotter, Joshua J Turner, Brian Kennedy, Franz Hennies, Frank MF de Groot, Kelly J Gaffney, Simone Techert, Michael Odelius, Alexander Föhlisch, Orbital-specific mapping of the ligand exchange dynamics of Fe(CO)$_5$ in solution, 2015, {\it Nature}, {\bf 520}, 78

\bibitem{Gent2017} William E. Gent, Kipil Lim, Yufeng Liang, Qinghao Li, Taylor Barnes, Sung-Jin Ahn, Kevin H. Stone, Mitchell McIntire, Jihyun Hong, Jay Hyok Song, Yiyang Li, Apurva Mehta, Stefano Ermon, Tolek Tyliszczak, David Kilcoyne, David Vine, Jin-Hwan Park, Seok-Kwang Doo, Michael F. Toney, Wanli Yang, David Prendergast and William C. Chueh,  2017, {\it Nature Communications} {\bf 8}, 2091 

\bibitem{Yang2018} Wanli Yang and Thomas Devereaux, Anionic and cationic redox and interfaces in batteries: Advances from soft x-ray absorption spectroscopy to resonant inelastic scattering, 2018, {\it Journal of Power Sources}, {\bf 389}, 188

\bibitem{FEFF} J.J. Rehr, J.J. Kas, F.D. Vila, M.P. Prange, K. Jorissen, Parameter-free calculations of x-ray spectra with FEFF9, {\it Phys. Chem. Chem. Phys.}, 2010, {\bf 12}, 5503

\bibitem{OCEAN} K. Gilmore, J. Vinson, E. L. Shirley, D. Prendergast, C. D. Pemmaraju, J. J. Kas, F. D. Vila, J. J. Rehr, Efficient implementation of core-excitation Bethe-Salpeter equation calculation., 2015, {\it Comp. Phys. Comm.} {\bf 197}, 109

\bibitem{JiaKedgeRIXS} CJ Jia, CC Chen, AP Sorini, B Moritz, TP Devereaux, Uncovering selective excitations using the resonant profile of indirect inelastic x-ray scattering in correlated materials: observing two-magnon scattering and relation to the dynamical structure factor, 2012, {\it New Journal of Physics}, {\bf 14}, 113038

\bibitem{JiaLedgeRIXS1} CJ Jia, EA Nowadnick, K Wohlfeld, YF Kung, C-C Chen, S Johnston, T Tohyama, B Moritz, TP Devereaux, Persistent spin excitations in doped antiferromagnets revealed by resonant inelastic light scattering, 2014, {\it Nature Comm.}, {\bf 5}, 3314

\bibitem{JiaLedgeRIXS2} Chunjing Jia, Krzysztof Wohlfeld, Yao Wang, Brian Moritz, Thomas P Devereaux, Using RIXS to uncover elementary charge and spin excitations, 2016, {\it Physical Review X}, {\bf 6}, 021020

\bibitem{Wannier90} AA Mostofi, JR Yates, G Pizzi, YS Lee, I Souza, D Vanderbilt, N Marzari, An updated version of wannier90: A tool for obtaining maximally-localised Wannier functions, 2014, {\it Comput. Phys. Commun.} {\bf 185}, 2309

\bibitem{Liang2017} Yufeng Liang and John T Vinson and Sri D Pemmaraju and Walter Drisdell and Eric L Shirley and David Prendergast, x-ray Absorption in Transition Metal Oxides: Self-Consistent Approaches versus a Many-Body Perturbation Theory, 2017, {\it Physical Review Letters} {\bf 118}, 096402

\bibitem{Liang2018} Yufeng Liang and David Prendergast, Quantum many-body effects in x-ray spectra efficiently computed using a basic graph algorithm, 2018, {\it Physical Review }B {\bf 20}, 205127

\bibitem{YangExperiment} Zengqing Zhuo, Chaitanya Das Pemmaraju, John Vinson, Chunjing Jia, Brian Moritz, Ilkyu Lee, Shawn Sallies, Qinghao Li, Jinpeng Wu, Kehua Dai, Yi-de Chuang, Zahid Hussain, Feng Pan, Thomas P. Devereaux, Wanli Yang, Spectroscopic Signature of the oxygen States in Peroxides, 2018, arXiv:1809.08292

\bibitem{QE} P. Giannozzi, S. Baroni, N. Bonini, M. Calandra, R. Car, C. Cavazzoni, D. Ceresoli, G. L. Chiarotti, M. Cococcioni, I. Dabo, A. Dal Corso, S. Fabris, G. Fratesi, S. de Gironcoli, R. Gebauer, U. Gerstmann, C. Gougoussis, A. Kokalj, M. Lazzeri, L. Martin-Samos, N. Marzari, F. Mauri, R. Mazzarello, S. Paolini, A. Pasquarello, L. Paulatto, C. Sbraccia, S. Scandolo, G. Sclauzero, A. P. Seitsonen, A. Smogunov, P. Umari, R. M. Wentzcovitch, , QUANTUM ESPRESSO: a modular and open-source software project for quantum simulations of materials, 2009, {\it Journal of Physics: Condensed Matter}, {\bf 21}, 39 

\end{thebibliography}
\end{document}